\documentclass{aa}
\usepackage[varg]{txfonts}
\usepackage{natbib}

\begin{document}

\title{Spectroscopy of BL Lac objects of extraordinary luminosity}

\author{M. Landoni\inst{1,2}
  \and R. Falomo\inst{3} 
     \and A. Treves\inst{4,2}
      \and B. Sbarufatti\inst{1}} 


\institute{INAF - Osservatorio Astronomico di Brera, Via Bianchi 46, I-23807 Merate (LC), Italy
\and INAF - Istituto Nazionale di Astrofisica and INFN - Istituto Nazionale di Fisica Nucleare
  \and INAF - Osservatorio Astronomico di Padova, Vicolo dell' Osservatorio 5, I-35122 Padova, Italy
  \and Universit\`{a}  degli Studi dell'Insubria. Via Valleggio 11, I-22100 Como, Italy. 
}

\date{Received 19 May 2014 / Accepted 11 July 2014}

\abstract {} {We aim to determine the redshift (or stringent lower limits) of a number of bright BL Lacs objects.} {We secured medium resolution optical and near-infrared spectra of 4 bright BL Lac objects of unknown redshift using the spectrograph X-Shooter at the ESO-VLT.} {In spite of the high quality of the spectra and the extended spectral range of the observations we have not detected intrisic spectral features for these sources. However we are able to provide strigent lower limits to their redshift. In particular, for the two TeV sources PG 1553+113 and H 1722+119 we infer $z > 0.30$ and $z > 0.35$ respectively. We also detect an intervening Ca II absorption doublet in the spectrum of MH 2136-428 that is ascribed to the the halo of a nearby giant elliptical galaxy at $\sim$ 100 kpc of projected distance.} {Under the hypothesis that all BL Lacs are hosted by luminous bulge dominated galaxies, the present state of art spectroscopic observations of bright BL Lacs indicate that these objects are likely sources with  extremely beamed nuclear emission .  We present simulations to show under which circustances it will be possible to probe this hypothesis from the detection of very weak  absorptions using the next generation of extremely large optical telescopes.} 

\maketitle

\section{Introduction}

BL Lac objects are active nuclei of massive elliptical galaxies characterized by a strong non thermal emission dominated by a relativistic jet that is closely aligned along the line of sight (see e.g. see Falomo Pian Treves Review A\&A 2014 in press). As a consequence, their optical spectra exhibit weak lines or a featureless continuum thus preventing the determination of their redshift.
Nevertheless, the estimate of distance of BL Lac objects is mandatory for constraining models of their emission, considering also that these objects are the dominant population of the extragalactic sky at high energies,  in particular for the TeV band (e.g. \citep{costa13}). 
In the absence of direct redshift measurement through the nuclear emission lines or absorptions from host galaxies, other procedures have been proposed. Some of them are based on the measurement of the absolute magnitude of the host galaxy, that can be assumed as a standard candle (see \citealt{sbahst} and references therein) while others exploit the detection of intervening absorption lines toward the source in order to put a lower limit to the redshift. For instance, this method was successfully exploited with HST/COS observations of  Lyman-$\alpha$ forest in FUV constraining the redshift of a number of bright BL Lacs at z $\sim$ 0.4 (see e.g. \cite{danfor10, furn13}).  Another approach to obtain redshift lower limits of BLL is to use their spectral energy distribution over a very large spectral range \citep{rau12}.  \\
\indent The improved S-N ratio of optical spectra collected through new generation large telescopes, combined with high throughput instruments, has also been successfully adopted in the determination of redshifts of BL Lacs (e.g. \cite{sba06, land13, shaw13} and references therein). The campaigns carried out by \cite{sba06} and \cite{land13} at the 10mt class telescopes VLT with FORS2 spectrograph, with an improvement of the S-N ratio of up to $\sim$ 300 for bright sources, allowed the determination of redshifts for many objects that had featureless spectra based on observations collected by 4mt class telescopes. \\
\indent In this context, the new-generation of spectrographs that combine high resolution and wide spectral coverage yield a considerable improvement on these issues. In this light, the state-of-the-art X-Shooter spectrograph at ESO-VLT is the most natural choice to tackle BL Lac spectroscopy. In fact, with this instrument it is possible to detect very faint spectral features  (EW $\sim$ 1 order of magnitude smaller than those detectable with previous instruments). This translates into the possibility to infer stringent lower limit to the redshift of the sources in which no spectral features (intrinsic or due to intervening systems) are present. \\
In order to exploit these new capabilities we obtained spectroscopy of five bright BL Lac objects (R  $\leq$ 17) which had defeated previous redshift determinations. They are therefore expected to be extremely luminous in the optical band or extremely beamed. In this work we report the results for four sources since PKS 0048-097 was already discussed in \cite{landoni12}. \\
\indent The paper is organised as follows. Our observations and data reduction are detailed in Section 2 while main spectroscopical results are reported in Section 3. Discussion and conclusions on the overall sample, with an outlook on future perspective, are given in Section 4.
Throughout the paper, we consider the following cosmological parameters $H_{0} = 70$ km s$^{-1}$    Mpc$^{-1}$, $\Omega_{m} = 0.27$, $\Omega_{\Lambda} = 0.73$.

\section{Observations, data reduction and analysis}
We secured UVB (3200 - 5500 $\textrm{\AA}$), VIS (5600 - 10.000 $\textrm{\AA}$) and NIR (11000 - 15.000 $\textrm{\AA}$) spectra of the targets in service mode using the European Southern Observatory (ESO) Very Large Telescope (VLT) equipped with the X-SHOOTER spectrograph \citep{vernet11}. The instrument was configured in standard NODDING mode adopting the recommended ABBA sequence for the exposures in order to guarantee an accurate sky subtraction in the extracted frames. The Journal of Observation is reported in Table \ref{table:jobs}. 
\\
Data reduction was performed by the X-Shooter Data Reduction Pipeline (version 2.0.0, see \citet{goldoni11}) in polynomial mode.
In particular, for each observation, we calculated the master bias frames and master flat frames for each channel. We also obtained the bidimensional mapping required by the pipeline stack to resample the echelle orders. Finally, for each channel, we computed the sensitivity functions to calibrate the spectra in flux using spectrophotometric standard stars. 
\\
In order to take into account slit loss effects and overall systematics on the flux calibration of the spectra we performed a photometry on the acquisition images of the targets, secured in R band. The overall uncertainty on the flux (based on the estimation of the S-N ratio inside the aperture considered for photometry) is $\Delta m \sim 0.20$ while the wavelength calibration is accurate within $\Delta \lambda \sim 0.5 \textrm{\AA}$. Checking the weather conditions at Paranal, the targets have been observed under clear sky condition with the exception of PG 1553+113 and, marginally, H 1722+119 due to thin clouds during the exposure. The acquisition images have been secured with clear sky.\\
\indent From the calibrated spectra of each source, we first compute the minimum detectable equivalent width following the procedure described in \cite{sba05} and we search for spectral features with EW greater than the minimum measurable one. 
In the case of absence of detectable features, we exploit the capability of the instrument to measure minimal EW in the range 4000-8000 $\textrm{\AA} $ in order to put stringent lower limit to sources which exhibit totally featureless spectra. We evaluate the $EW_{min}$ on the wavelength range 4000-8000 $\textrm{\AA}$ since it corresponds to a redshift interval of about $0.01 \leq z \leq 1.00$ for the most prominent absorption lines of the host galaxy. This choice appears reasonable because if the object lies at redshift higher than $z \geq 1$ the detection of the underling host galaxy is challenging even with medium resolution spectroscopy with reasonable S-N ratio ratio for 8mt class telescopes, while even higher redshifts are also ruled out by the absence of Ly-$\alpha$ forest in the UV part of the X-SHOOTER spectrum. 
The minimum EW on the spectra,  the apparent magnitude of the nucleus and the redshift are related through the equation 
\begin{equation}
\textrm{EW}_{\textrm{obs}} = \frac{(1+z)\textrm{EW}_0}{1+\rho_0\Delta(\lambda)A(z)}
\end{equation}
where EW$_{obs}$ is the observed minimum equivalent width, EW$_0$ is the equivalent width of the feature in the host galaxy template of Kinney et al. (1996), $\rho_0$ is the nucleus-to-host flux ratio in the rest frame R band effective wavelength and $A(z)$ is the aperture correction. The term $\Delta(\lambda)$ is a normalisation factor which takes into account the redshift of the source.
The knowledge of $\textrm{EW}_{\textrm{obs}}$, which is the minimum detectable one, and the apparent magnitude of the nucleus allow to apply Equation (1) to derive a lower limit to the redshift (details are given in \cite{sba06} and references therein). We applied this procedure to our four spectra reporting the results in Table \ref{table:spec}.\\
\indent Finally, in order to take the advantage of the available extended spectral range, we performed a synthetic narrow band photometry (integrating the flux density in various bins of 200$\AA$) on the spectrum of each object aimed to construct the Spectral Energy Distribution (SED).

\begin{table*}
\caption{Journal of observations}              
\label{table:jobs}      
\centering     
\begin{tabular}{c c c c c c c  l}
\hline\hline 
  Object \tablefootmark{a}  & Date of observation \tablefootmark{b} & Seeing \tablefootmark{c} &  Slit width \tablefootmark{d}& R \tablefootmark{e}& $t_{exp}$ \tablefootmark{f} & $N$ \tablefootmark{g} & Channel \tablefootmark{h}  \\

  \hline
  
  & &   & $1.6^{\prime\prime} \times 11^{\prime\prime}$ & 3300 & 2720 & 4  &(UVB)\\

PG 1553+113 & 28 Apr 2010 & 1.30  & $1.5^{\prime\prime} \times 11^{\prime\prime}$ & 5400  & 2460 & 6  &(VIS)\\
 & &  &$1.5^{\prime\prime} \times 11^{\prime\prime}$ & 3500  & 1440 & 6& (NIR)\\
  \hline
  
    & &  & $1.6^{\prime\prime} \times 11^{\prime\prime}$ & 3300 & 2720 & 4  &(UVB)\\

H 1722+119 & 28 Apr 2010 & 1.60  & $1.5^{\prime\prime} \times 11^{\prime\prime}$ & 5400  & 2460 & 6 & (VIS)\\
 & &  &$1.5^{\prime\prime} \times 11^{\prime\prime}$ & 3500  & 1440 & 6 &(NIR)\\
  \hline
  
      & &   & $1.6^{\prime\prime} \times 11^{\prime\prime}$ & 3300 & 2720 & 4 & (UVB)\\

MH 2136-428 & 28 Apr 2010 & 1.40  & $1.5^{\prime\prime} \times 11^{\prime\prime}$ & 5400  & 2460 & 6 &(VIS)\\
 & &  &$1.5^{\prime\prime} \times 11^{\prime\prime}$ & 3500  & 1440 & 6 & (NIR)\\

   \hline

 & &   & $1.6^{\prime\prime} \times 11^{\prime\prime}$ & 3300  & 5440 & 4 & (UVB) \\
PKS 2254-204 & 30 Ago / 06 Sep 2010 &1.20  & $1.5^{\prime\prime} \times 11^{\prime\prime}$ & 5400  & 4920 & 6 & (VIS)\\
 & &  & $1.5^{\prime\prime} \times 11^{\prime\prime}$ & 3500  & 2880 & 6  &(NIR)\\
  \hline
  
\end{tabular}
\tablefoot{
\tablefoottext{a}{IAU name}
\tablefoottext{b}{Date of observation}
\tablefoottext{c}{Seeing during the observation (measured on the acquisition images).}
\tablefoottext{d}{Slit width (in arcsec)}
\tablefoottext{e}{Average Resolution ($\lambda / \delta\lambda$)}
\tablefoottext{f}{Total integration time (in seconds).}
\tablefoottext{g}{Number of spectra obtained in nodding mode.
\tablefoottext{h}{X-Shooter Channel}
}
}

\end{table*}
\section{Results}
\begin{center}
\begin{figure*}
   \includegraphics[width=18cm]{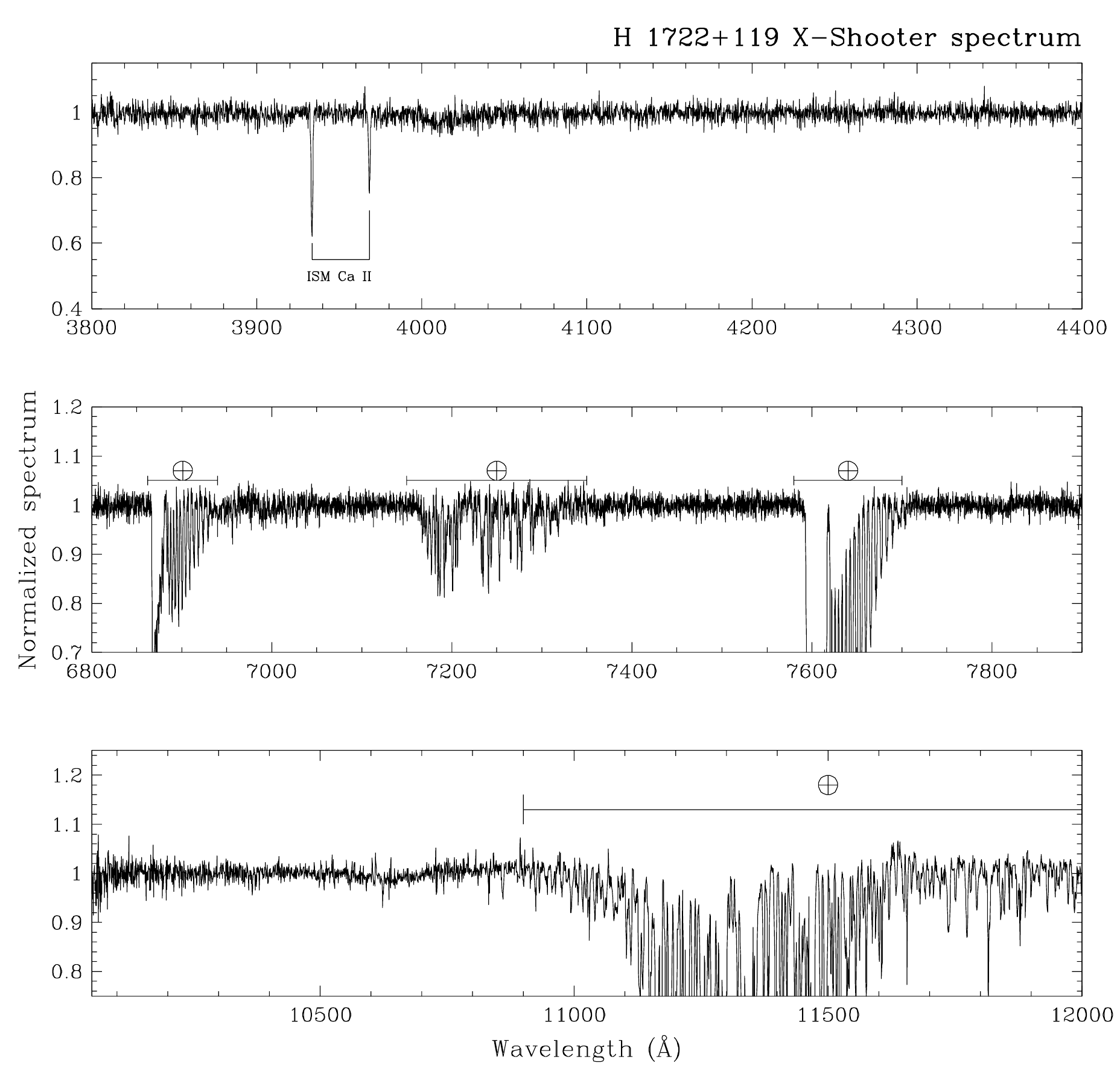}
     \caption{H 1722+119 X-SHOOTER normalised spectrum in three sample ranges (first panel: UVB channel; middle panel: VIS channel; bottom panel: near-IR channel). With the exception of galactic and telluric features, no intrinsic or intervening spectral lines are detected in the full range $\sim$0.33 - 1.5 $\mu$m}
     \label{fig:samplespec}
\end{figure*}
\end{center}
\begin{center}
\begin{figure*}
   \includegraphics[width=18cm]{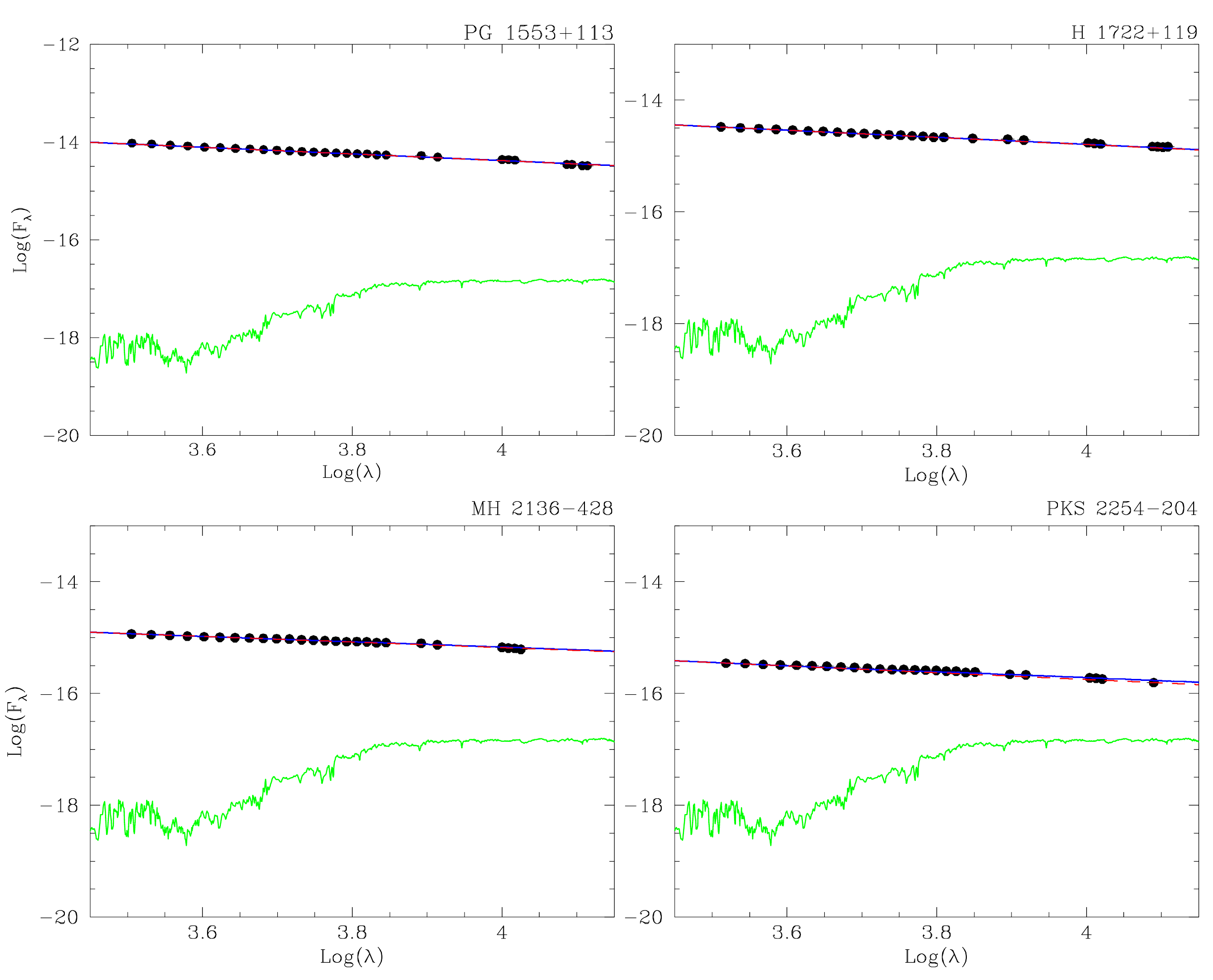}
     \caption{SEDs of the four bright BL Lac object obtained from X-SHOOTER spectra in the wavelength range 3200-15000 $\textrm{\AA}$. The blue solid line is the sum of the two components:  the non-thermal emission which is described by a single power law (red dashed line) and the contribution from a luminous host galaxy starlight ($M_r = -23.90$) at $z = 0.50$  (green solid line). The superposition of the blue and red lines indicate that such a galaxy is hidden by the very bright nuclear emission  }
     \label{fig:seds}
\end{figure*}
\end{center}

We report in Figure \ref{fig:samplespec} an example of the spectra in three different ranges from optical to near-IR. The full spectra for the 4 sources are available both in the electronic edition of the paper and on the website database \texttt{http://archive.oapd.inaf.it/zbllac/}. 
With the exception of MH 2136-428 where we are able to detect an intervening Ca II system at $z \sim 0.008$ (see Section Notes on individual sources), the other spectra do not show features apart from the galactic and telluric ones. 
The lower limit to the redshift  for each source are also reported in Table \ref{table:spec}. \\
\indent In Figure \ref{fig:seds} we show the SEDs obtained with the procedure described in Section 2. The observed emission is the sum of two components: a non thermal power which arises from the relativistic jet (single power law) and host galaxy starlight contribution at z $\sim 0.5$ from a luminous giant elliptical of $M_r = -23.90$, which is $\sim$ 1 mags more luminous than a typical BL Lac host galaxy (giant elliptical of $M_r = -22.90$, \cite{sbahst}). As illustrated in Figure \ref{fig:seds}, the emission of each source is completely dominated by the non thermal component and the starlight contribution from the galaxy, despite its luminosity, is negligible.

\begin{table*}

\caption{Redshift lower limits and properties of observed sources.}              
\label{table:spec}      
\centering     
\begin{tabular}{l l l l l l l l l l l l | | l| |l | |l |}
\hline\hline 
  Object \tablefootmark{(a)}  & R\tablefootmark{(b)} & EW$_{min}$ \tablefootmark{(c)} & S-N Ratio \tablefootmark{(d)}  & $\alpha$ \tablefootmark{(e)} & M$_{R}$\tablefootmark{(f)} & N-H ratio \tablefootmark{(g)} & $z$ \tablefootmark{(h)} & $\delta_{opt}$\tablefootmark{(k)} \\
  \hline

\object{PG 1553+113} & 13.80 & 0.04 & 140 $\pm$ 30 & 0.90 &$\le$ -27.2 & 50 & $\geq$ 0.30 & $\geq$ 700\\ 
\object{H 1722+119} & 14.90 & 0.07 & 80 $\pm$ 20 & 1.00 & $\le$ -26.3 & 25 & $\geq$ 0.35 & $\geq$ 400\\ 
\object{MH 2136-428} & 15.80 & 0.06 & 100 $\pm$ 30 & 0.75 & $\le$ -26.0 & 15 & $\geq$ 0.50 & $\geq$ 500\\ 
\object{PKS 2254-204} & 17.10 & 0.10 & 60 $\pm$ 10 & 0.75 & $\le$ -25.4 & 10 & $\geq$ 0.65 & $\geq$ 300\\

\hline
\end{tabular}
\tablefoot{
\tablefoottext{a}{IAU name}
\tablefoottext{b}{R band apparent magnitude measured from our spectrum.}
\tablefoottext{c}{Average EW min in the range 4000-8000$\textrm{\AA}$.}
\tablefoottext{d}{Average S-N ratio in the range 4000-8000$\textrm{\AA}$.}
\tablefoottext{e}{Spectral index.}
\tablefoottext{f}{Nucleus absolute magnitude limit in R band computed assuming lower limit on the z.}
\tablefoottext{g}{Nucleus to host ratio lower limit assuming BL Lac host galaxy standard candle with $M_r = -22.90$}
\tablefoottext{h}{Redshift lower limit}
\tablefoottext{k}{Optical beaming factor computed following the procedure described in \cite{land13}.}
}

\end{table*}

\subsection{Notes on individual sources}
\begin{center}
\begin{figure}
   \includegraphics[width=8cm]{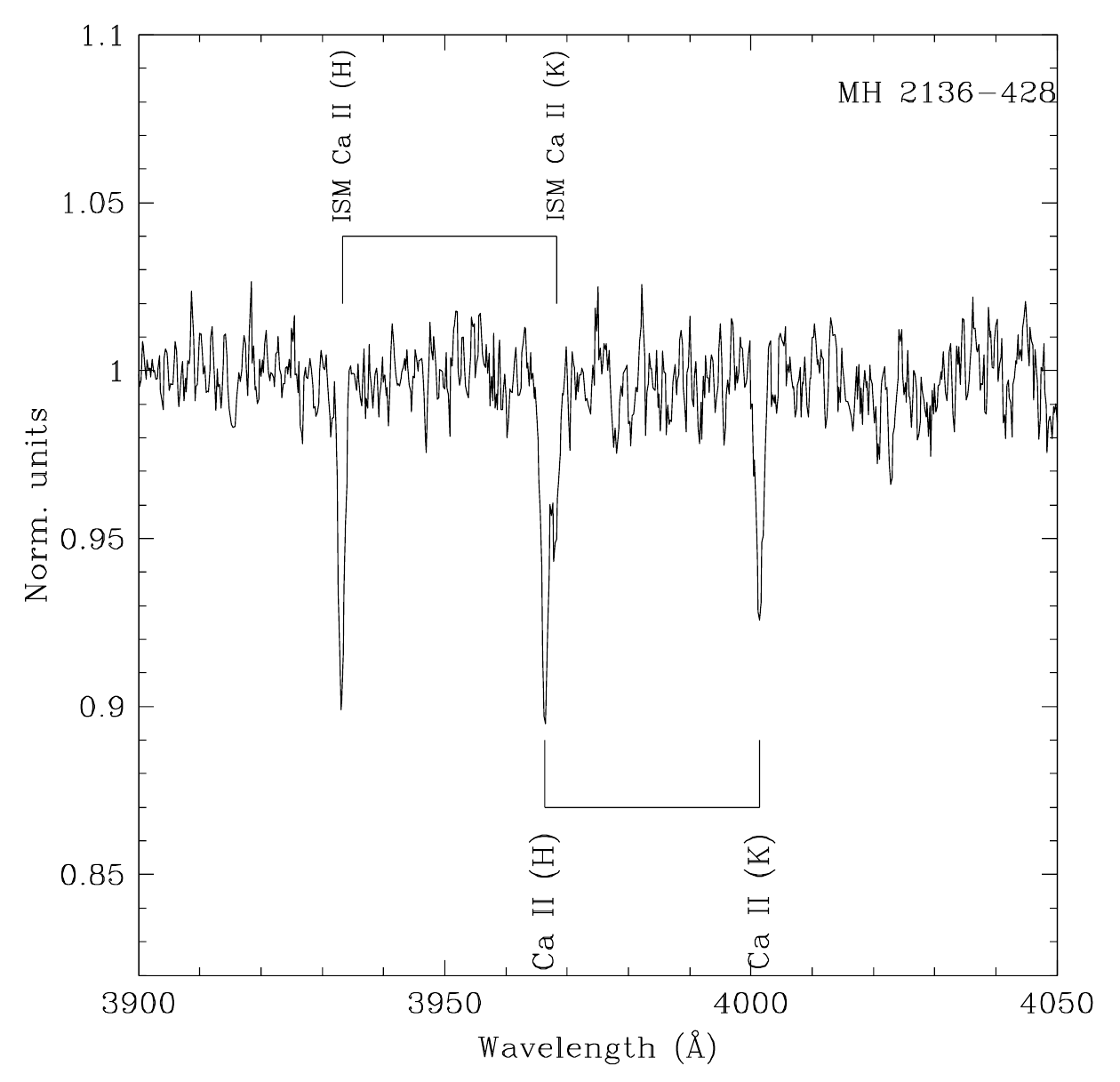}
     \caption{MH 2136-428 Ca II intervening system arising in the nearby galaxy NGC 7097.}
     \label{fig:2136-deb}
\end{figure}
\end{center}

\begin{center}
\begin{figure*}
   \includegraphics[width=17cm]{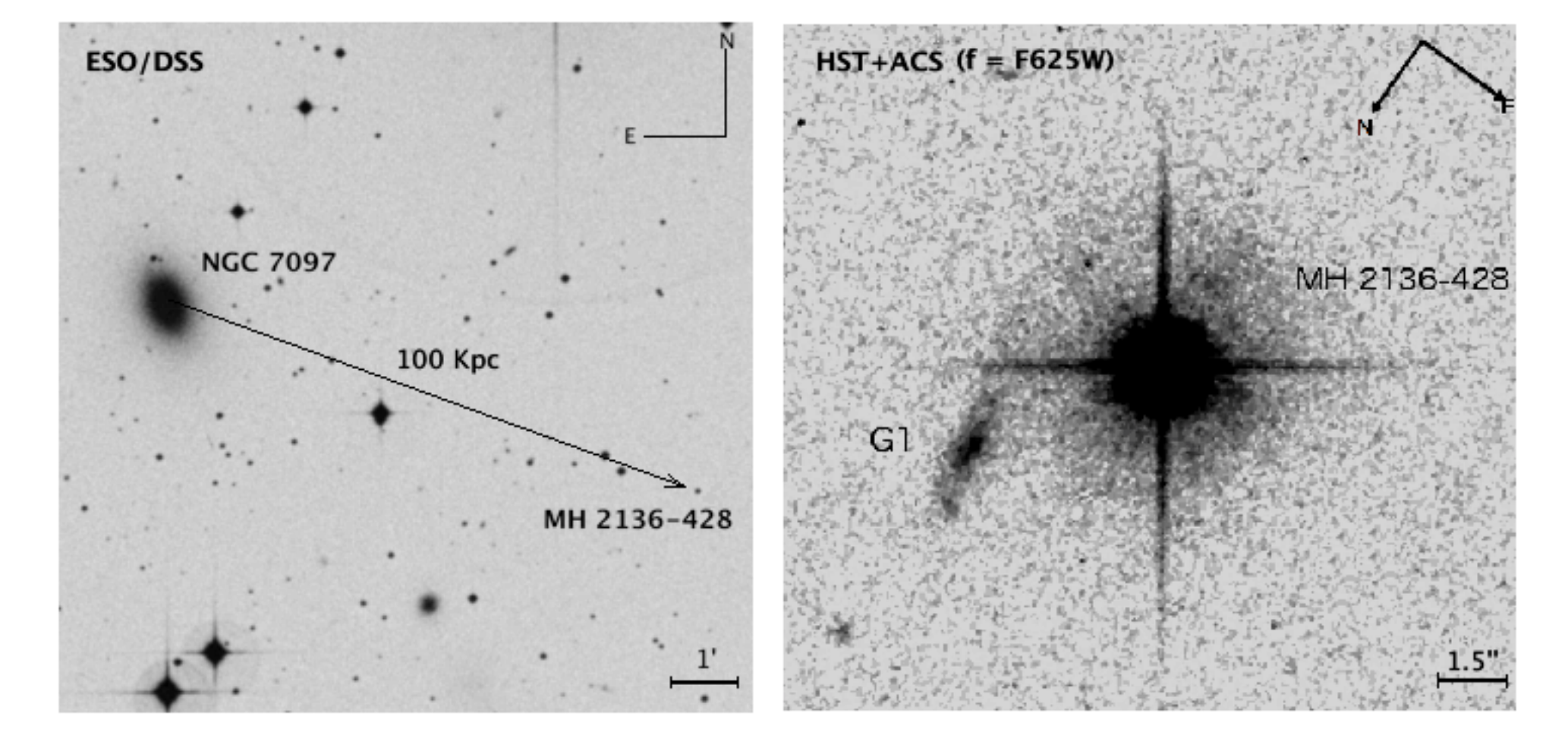}
     \caption{The BL Lac object MH 2136-428. Left Panel:ESO/DSS field that shows the galaxy NGC 7097 possible associated to the Ca II absorption system. Right panel: HST+ACS image (integration time is 780s; f = F625W). The image clearly shows the presence of a faint galaxy at 3.2$^{\prime\prime}$ NW from the BL Lac (labelled as G1). The target is unresolved and the faint galaxy (G1) at $\sim$ 3.2$^{\prime\prime}$ NW is likely a background object.}
     \label{fig:2136-hst}
\end{figure*}
\end{center}

\textbf{PKS 1553+113}: This very bright (R $\sim$ 14) BL Lac source, that appeared in the Palomar-Green catalog  of ultraviolet-excess stellar objects \citep{palomgreen}, has been thoroughly studied from radio band to VHE (we mention the recent availability of Cherenkov telescopes observations from e.g. \cite{prandini10,magic12}). In the near UV and optical bands (see e.g. \cite{fal90, fal93}) its spectrum appears to be featureless and dominated by the non thermal emission. Spectroscopy with 8m class telescope at VLT \citep{sba06} also confirmed featureless nature of the spectrum and derived a lower limit for the source of $z > 0.1$ based on the lack of absorption features from the starlight component. The non-detection of a standard host galaxy in the HST images \citep{treves07} allowed to estimate z $\geq$ 0.30-0.40 making it one of the most distant TeV extragalactic source. These limits are consistent with two recent constraints on the redshift. The first one is obtained through far UV spectroscopy with HST/COS ($z \sim 0.40 - 0.50$, see \cite{danfor10}) exploiting Lyman-$\alpha$ absorption systems toward the line of sight of the source, while the other is based on the observed absorption of the high energy TeV due to photon-photon interaction with the Extragalactic Background Light , yielding $z \leq 0.60$ (see \cite{prandini10}). The X-SHOOTER spectroscopy presented in this paper confirms the featureless nature of the spectrum on the wide spectral range 3200 - 15000 $\textrm{\AA}$. Exploiting the capability to measure lines with very small equivalent width we calculated a new redshift lower limit obtaining $z \geq 0.30$.
\\
\\
\textbf{H 1722+119}: This source (R $\sim$ 15 ) was studied by \cite{allen82} through spectrophotometry carried out with the 4m class Anglo-Australian telescope in late 80's confirming its \textit{bona-fide} BL Lac nature \citep{allen82}. This object is also a well known strong radio \citep{hea07}, X-ray \citep{donato05} and $\gamma$-ray emitter (see e.g. \cite{abdo10}). The source has been recently discovered during a flare by MAGIC collaboration \citep{cortina13} in the TeV band. Previous spectroscopical study with 8m class telescope \citep{sba06} confirmed that H 1722+119 exhibits a completely featureless spectrum and put this object at $z \geq 0.17$. We infer from the minimum EW$_{min}$ measured on the spectrum a tighter lower limit to the redshift of $z > 0.35$. We also remark that the source is unresolved in the image obtained with HST/WFPC3 \citep{urryscarpa}.\\
\\
\textbf{MH 2136-428}: This is a bright BL Lac object $(\textrm{R} \sim 15)$ discovered through strong optical variability ($\Delta m \sim 1.0$) from the optical survey carried with the UK 1.2 Schmidt telescope \citep{haw91}. Spectroscopic studies (\cite{haw91}, \cite{sba06} and \cite{shaw13}) failed to detect intrinsic or intervening features. In our spectrum there are no significant intrinsic emission or absorption lines down to EW$_{\textrm{min}}$ 0.06 that set a lower limit of the redshift $z \geq 0.50$ (see Section 2). We note the presence of an intervening absorption system due to Ca II at $\lambda\lambda$ 3966.40 4001.20 (see the extended Figure 1 in electronic version and Figure \ref{fig:2136-deb}) corresponding to a radial velocity of 2450 $\pm$ 50 km s$^{-1}$. Note that the H line is blended with the ISM Ca II K line. The EWs of the intervening absorption (obtained from deblending of the line for the H band) are EW $0.15  \pm 0.05$ $\textrm{\AA}$ and $0.10 \pm 0.02$$\textrm{\AA}$. We searched for low redshift galaxies in the field of MH 2136-428 associated with the absorption. The most plausible object responsible for this absorption is the halo of the early type NGC 7097 ($v_{r} \sim 2400-2600$ km s$^{-1}$) at $\sim$ 100 Kpc from the target (see Figure \ref{fig:2136-hst}). 
The target was observed by HST/ACS (Proposal ID:9494, PI: Paolo Padovani ) in a program for studying jet features in blazars. The unpublished image (see Figure \ref{fig:2136-hst}) shows indeed the presence of a faint ($m_{G1} \sim 21$) spiral galaxy at $\sim$ 3$^{\prime\prime}$ North-West from the BL Lac. The faintness of G1 suggests that it is a background galaxy. We used the HST image to search for emission of host galaxy of the BL Lac. The analysis performed with AIDA \citep{uslenghi11} shows that the target is unresolved, consistently with the relatively high redshift deduced from our spectroscopy. 
\\

\noindent \textbf{PKS 2254-204}: This object (R $\sim$ 17) is a strongly polarised \citep{impey90} featureless BL Lac \citep{veronveron}. More recent high quality spectra confirm the featureless emission (see \cite{sba06}). Our observation allows us to further search for very faint features (down to EW $\sim$ 0.10 $\textrm{\AA}$) over a wider spectral range. In spite of this improvement the spectrum remains lineless indicating a rather high nuclear to host galaxy emission. Based on this $\textrm{EW}_{\textrm{lim}}$ we set a redshift limit of $z > 0.65$.

\newpage
\section{Discussion and conclusions}
\begin{center}
\begin{figure*}

   \includegraphics[width=18.50cm]{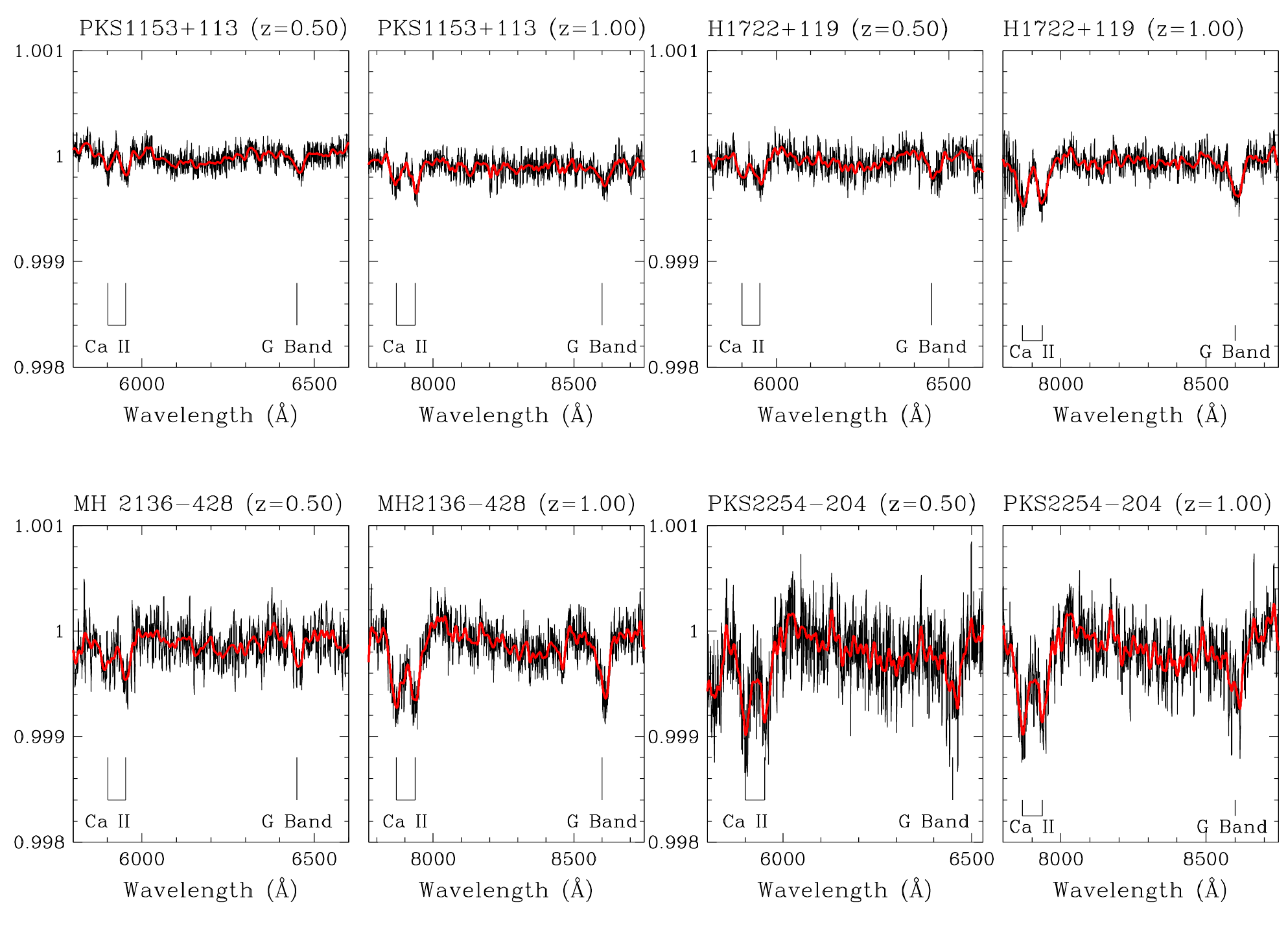}
     \caption{Simulated spectra of four BL Lacs using E-ELT equipped with an X-SHOOTER like instrument ($t_{exp} = 3600 s$, $R \sim 3000$). For each object we report two cases at $z = 0.50$ on the left panel and $z = 1.00$ on the right panel. The red solid line is the smoothed spectrum obtained by the adoption of a boxcar filter (9 pixels). The spectra represent the maximum N-H flux ratio for which it is possible to detect Ca II and G Band absorption lines of the host galaxy (see details in Table \ref{table:spec}).}
     \label{fig:simelt}
\end{figure*}
\end{center}

\begin{table*}
\caption{E-ELT X-SHOOTER like spectrograph simulations results.}              
\label{table:sim-res}      
\centering     
\begin{tabular}{l c| c c c c c| c c c c c}
\hline
\hline
 Source\tablefootmark{a} & m$_{nuc}$ \tablefootmark{b}& && $z = 0.50$ &  &&  & &$z = 1.00$   & \\
 \hline
   &  & SNR\tablefootmark{c}  & Obs. NH\tablefootmark{d} &  Int. NH\tablefootmark{e}  & m$_{host}$\tablefootmark{f}  & M$_{R}$\tablefootmark{g}   & SNR  & Obs. NH  &  Int. NH  & m$_{host}$ & M$_{R}$  \\
\hline
PG 1553+113 & 13.80 & 5300 & 2000 & 1000 & 21.30 & -21.40 & 6500 & 2500 & 1700 & 21.80 & -23.30\\

H 1722+119 & 14.90 & 3200 & 700 & 350 & 21.25 & -21.30 & 4000 & 1200 & 800 & 22.20 & -22.70 \\ 
MH 2136-428 & 15.80 & 2000 & 500 & 250 & 21.80 & -20.80 & 2500 & 800 & 540 & 22.90 & -22.10 \\ 
PKS 2254-204 & 17.10 & 1000 & 300 & 150 & 22-60 & -20.10 & 1300 & 600 & 400 & 23.60 & -21.40 \\
  \hline

  \hline

\end{tabular}
\tablefoot{
\tablefoottext{a}{Source simulated with E-ELT X-SHOOTER like spectrograph.}
\tablefoottext{b}{Apparent nucleus magnitude in R band.}
\tablefoottext{c}{Obtained Signal-to-Noise ratio for 3600s of exposure.}
\tablefoottext{d}{Observed Nucleus-to-host ratio at Ca II H\&K band host galaxy features.}
\tablefoottext{e}{Integral Nucleus-to-host ratio (assuming a slit loss for an aperture of 3$^{\prime\prime} diameter)$.}
\tablefoottext{f}{Apparent host magnitude in R band.}
\tablefoottext{g}{Absolute R band host magnitude.}
}

\end{table*}
\noindent We obtained high quality spectra from near-UV to near-IR of a small sample of very bright BL Lac objects of unknown redshift. In spite of the significant improvement in terms of S-N ratio and spectral resolution of the new spectra with respect to previous data,  we do not detect any intrinsic spectral feature arising from these sources. 
From the lack of starlight absorptions we derived stringent lower limits to their redshift that, combined with the brightness of the objects, indicates that these BL Lacs have a very high nuclear to host ratio (see Table \ref{table:spec}) compared to the average NHR values (0.1 to 1.0) as derived from \cite{urryscarpa,scarpahst}. The extreme nucleus absolute magnitude and N-H ratio (in particular for PG 1553+113 and H 1722+119) may be ascribed to an intrinsic huge luminosity, to a major beaming, or to both effects combined togheter. We notice that, assuming the presence of underlying broad emission lines one can estimate an optical beaming factor $\delta_{opt}$ which represents the ratio of the non-thermal (jet) to thermal component (disk) of the accreting nucleus. Following \cite{land13} and supposing a value for $z \sim 0.40$ such that Mg II ($\lambda_0 = 2800)$ is in the region $4000-8000 \textrm{\AA}$ one can calculate the lower limit to $\delta_{opt}$ by adopting the minimum observable EW$_{min}$ (see Table \ref{table:spec}). Under this hypothesis, according to values reported in Table \ref{table:spec}, the presence of an extreme beaming is suggested.\\
\indent Although the main result of the paper is to fix lower limit to the $z$ of the objects, one can put also upper limits exploiting the incidence of Mg II absorption systems. Considering the distribution of those systems with EW $\geq 0.2 \textrm{\AA}$ \citep{zhu13} we evaluate that the expected number of system per source should to be $\sim$ 0.8. In the spectra of our 4 sources we do not detect absorption systems of Mg II. Combining this fact with the TeV detection of 2 sources (PG 1553+113 and H 1722+119) we infer that, in average, the upper limit to the redshift should be $z \lesssim 0.50$.\\
\indent The observations presented in this paper highlight the today capabilities to explore the spectroscopic properties of BL Lac objects and thus the determination of their redshift or tighter lower limits. In order to make a significant improvement in this field a major contribution will come from the next generation of extremely large (30-50m class) telescopes, such as the Thirty Meter Telescope\footnote{\texttt{http://www.tmt.org}} and European Extremely Large Telescope (E-ELT)\footnote{\texttt{http://www.eso.org/sci/facilities/eelt/}}. In fact with these new facilities it will become possible to obtain very high S-N ratio spectra of bright BL Lacs and therefore detect the absorption features of their host galaxies under extreme conditions. 
To exemplify this point we performed  simulations of the spectra of our targets as could be secured by an X-SHOOTER like instrument (R $\sim$ 3000) coupled to the E-ELT including moderate adaptive optics facility. For the simulation we assume a plate scale of 0.03 $^{\prime\prime}$ px$^{-1}$ and a PSF with $\sim$ 90\% of Encircled Energy within an aperture of 3$^{\prime\prime}$ in order to maximise the host galaxy flux inside the aperture. We estimated the S-N ratio of each object using the AETC\footnote{ \texttt{http://aetc.oapd.inaf.it}} \citep{falomoaetc}. We produced simulated spectra of the targets, assuming $z = 0.50$ and $z = 1.00$, in order to evaluate the maximum nucleus to host galaxy flux ratio that allows one to detect the absorption lines of the galaxies (see Figure \ref{fig:simelt}). Result of the simulations are summarised in Table \ref{table:sim-res}. It turns out that with these future facilities it will be possible to derive the redshift of these objects under extreme conditions of observed N-H ratios: from $\sim 300$ to $\sim 2500$ depending on redshift and target brightness. This correspond to total N-H ratios in the range 100-1700 that is 1-2 order of magnitude higher than the average found for resolved BL Lacs \citep{urryscarpa}.

\begin{appendix}
\begin{center}
\begin{figure*}
   \includegraphics[width=18cm]{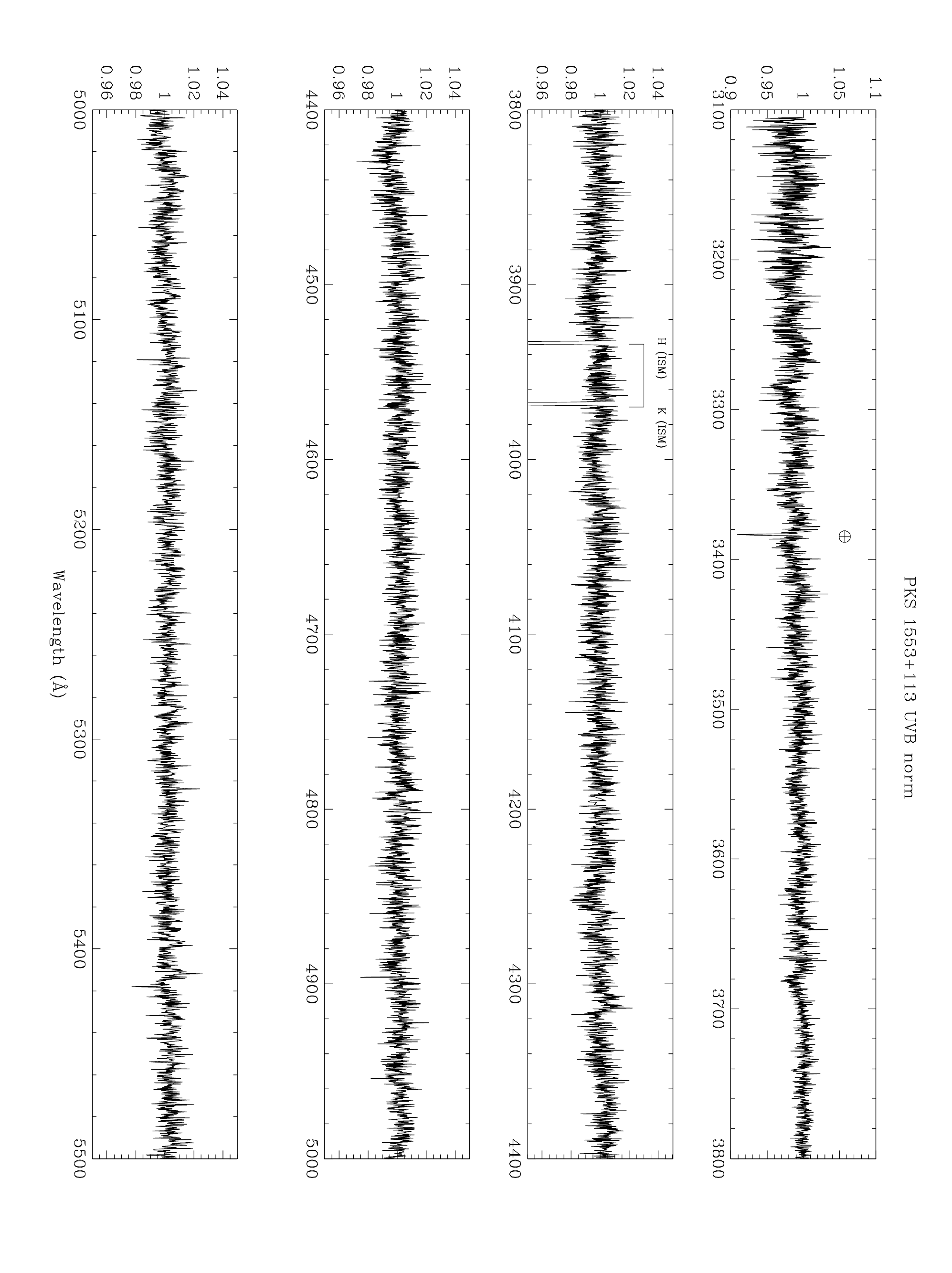}
     \caption{(continued) PG 1553+113 X-SHOOTER normalized UVB spectrum.}
     \label{fig:1553-uvbn}
\end{figure*}
\end{center}

\setcounter{figure}{0}
\begin{center}
\begin{figure*}
   \includegraphics[width=18cm]{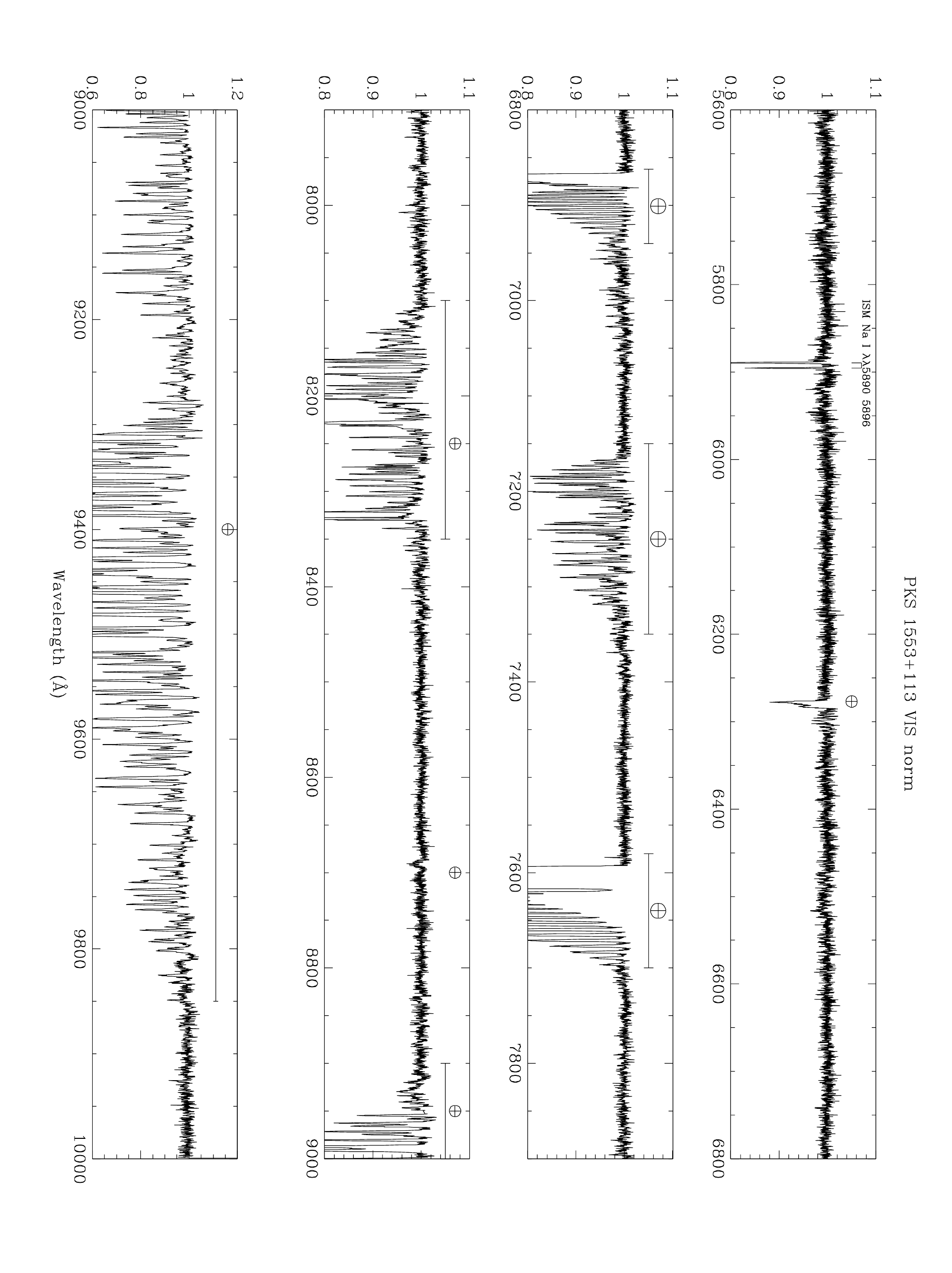}
     \caption{PG 1553+113 X-SHOOTER normalized optical spectrum.}
     \label{fig:1553-visn}
\end{figure*}
\end{center}
\setcounter{figure}{0}
\begin{center}
\begin{figure*}
   \includegraphics[width=18cm]{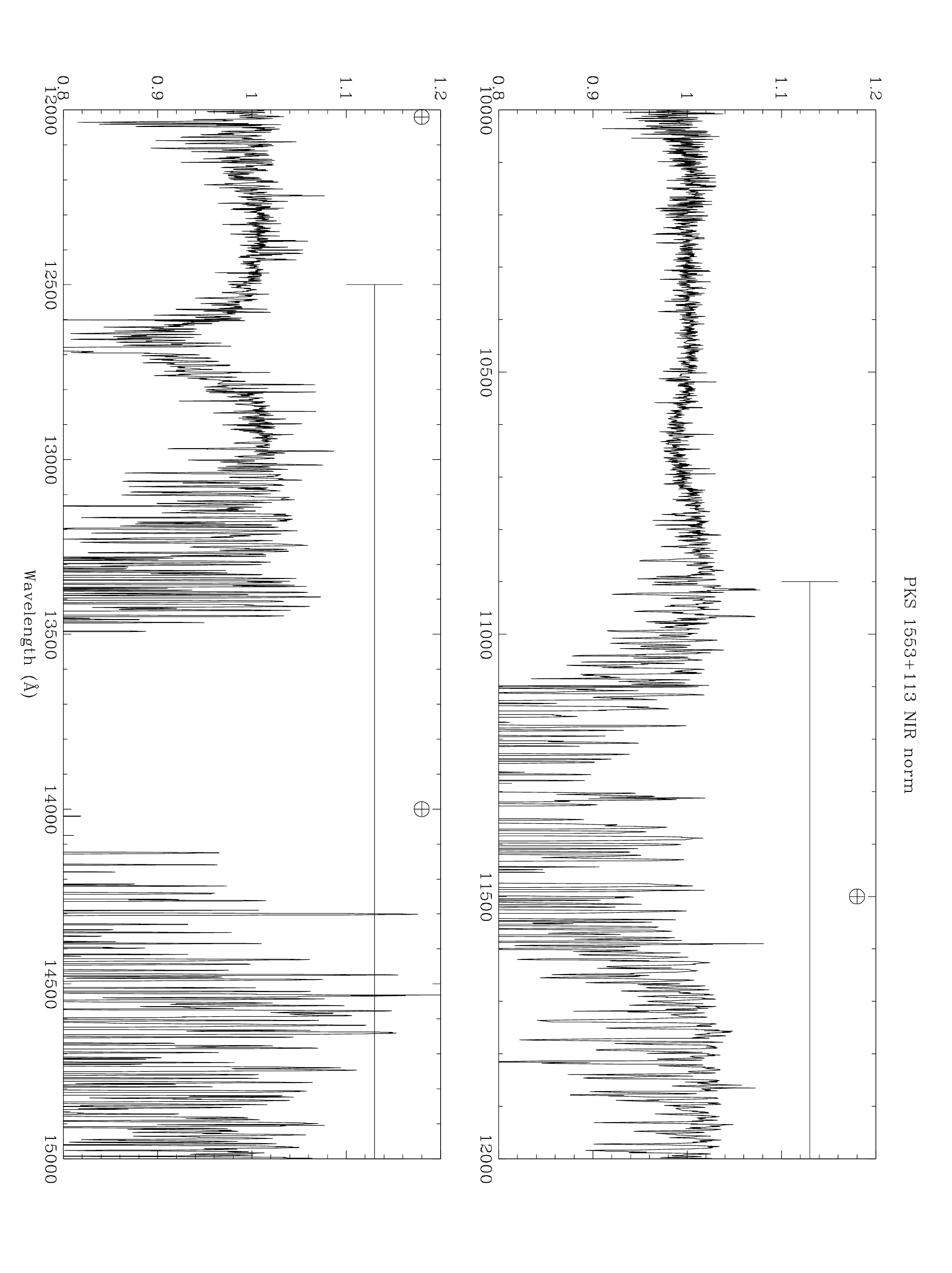}
     \caption{PG 1553+113 X-SHOOTER normalized near-IR spectrum.}
     \label{fig:1553-nirn}
\end{figure*}
\end{center}

\setcounter{figure}{0}
\begin{center}
\begin{figure*}
   \includegraphics[width=18cm]{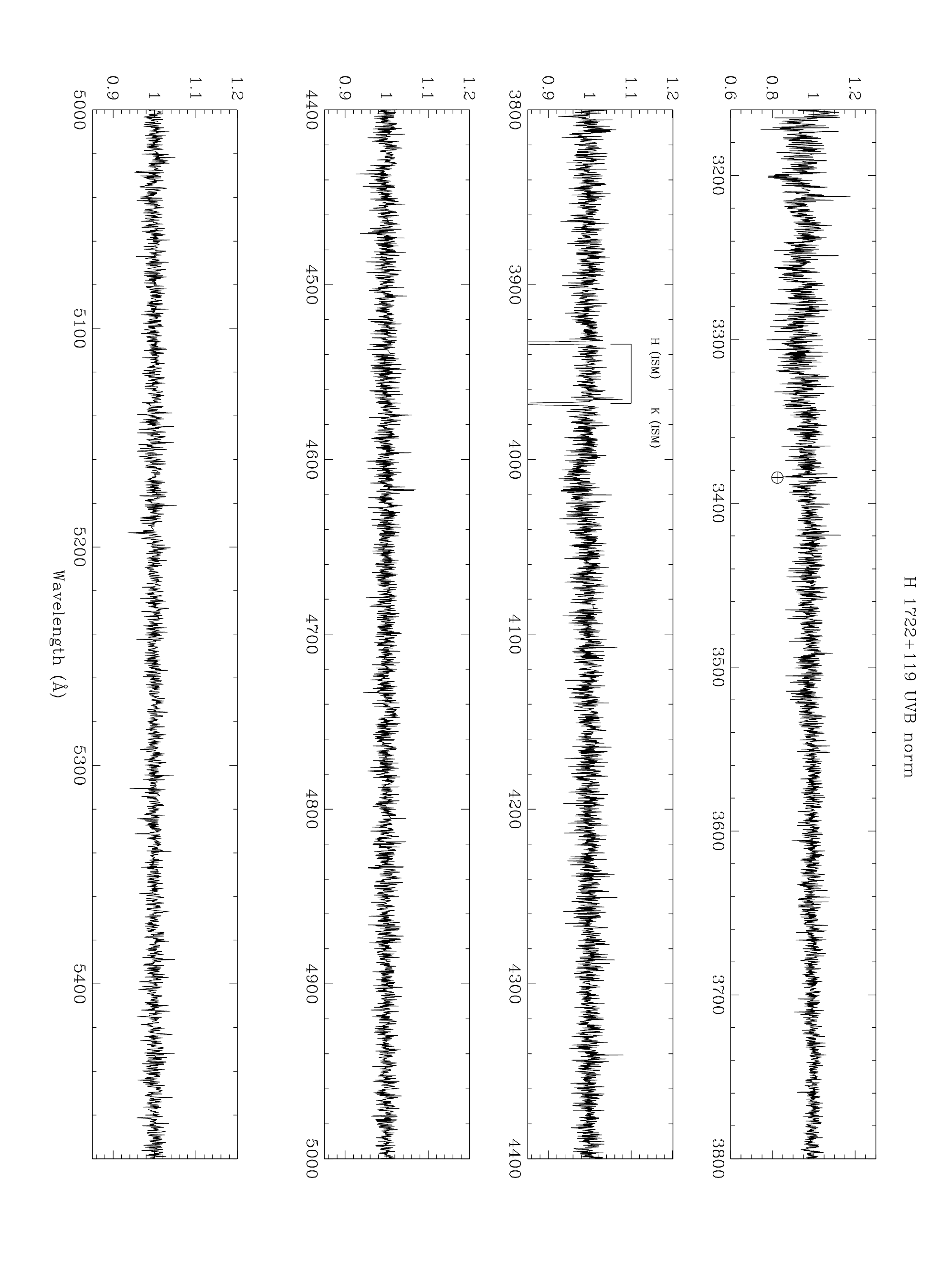}
     \caption{H 1722+119 X-SHOOTER normalized UVB spectrum.}
     \label{fig:1722-uvbn}
\end{figure*}
\end{center}

\setcounter{figure}{0}
\begin{center}
\begin{figure*}
   \includegraphics[width=18cm]{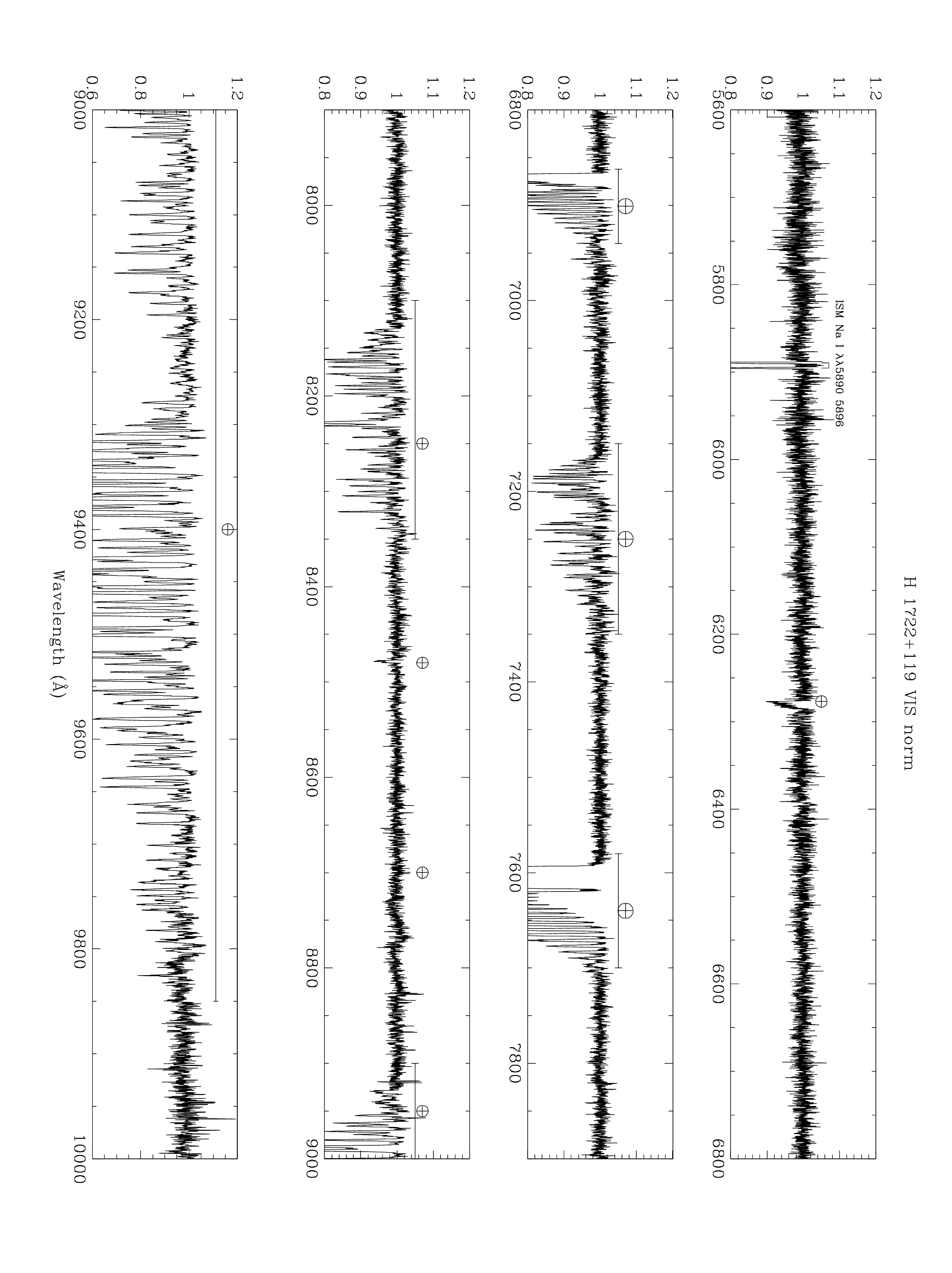}
     \caption{H 1722+119 X-SHOOTER normalized optical spectrum.}
     \label{fig:1722-visn}
\end{figure*}
\end{center}

\setcounter{figure}{0}
\begin{center}
\begin{figure*}
   \includegraphics[width=18cm]{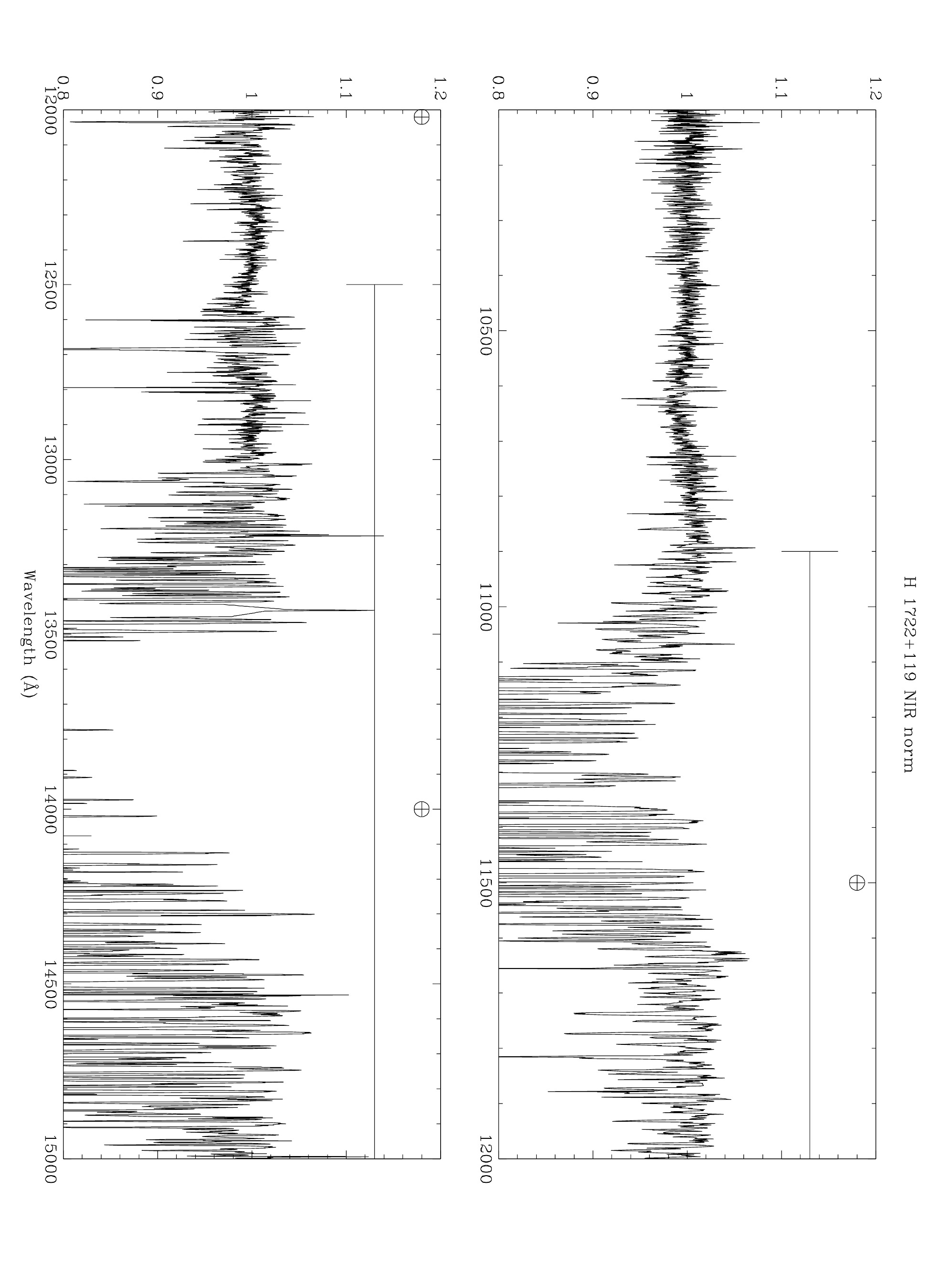}
     \caption{H 1722+119 X-SHOOTER normalized near-IR spectrum.}
     \label{fig:1722-nirn}
\end{figure*}
\end{center}

\setcounter{figure}{0}
\begin{center}
\begin{figure*}
   \includegraphics[width=18cm]{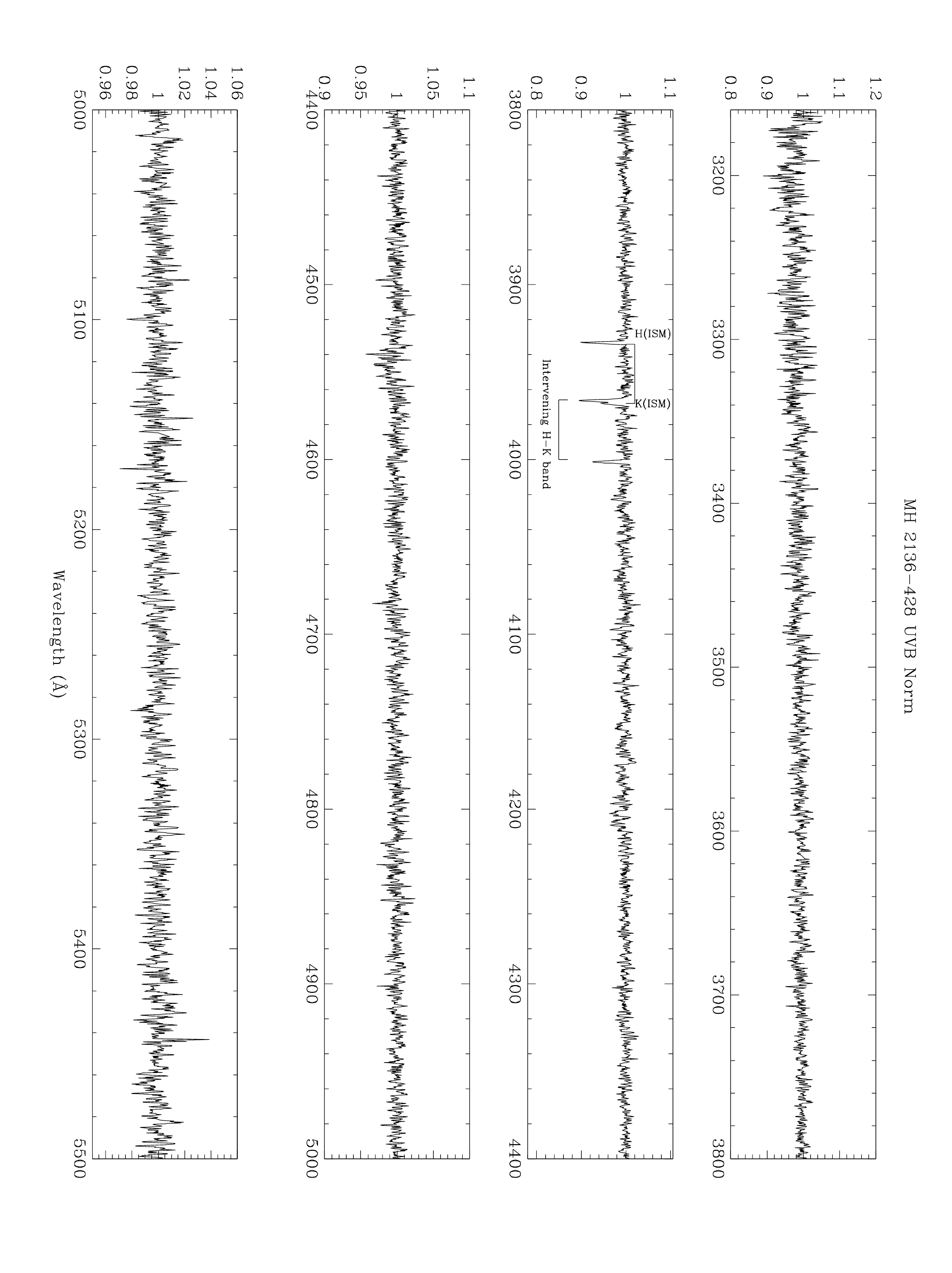}
     \caption{MH 2136-428 X-SHOOTER normalized UVB spectrum.}
     \label{fig:2136-uvbn}
\end{figure*}
\end{center}

\setcounter{figure}{0}
\begin{center}
\begin{figure*}
   \includegraphics[width=18cm]{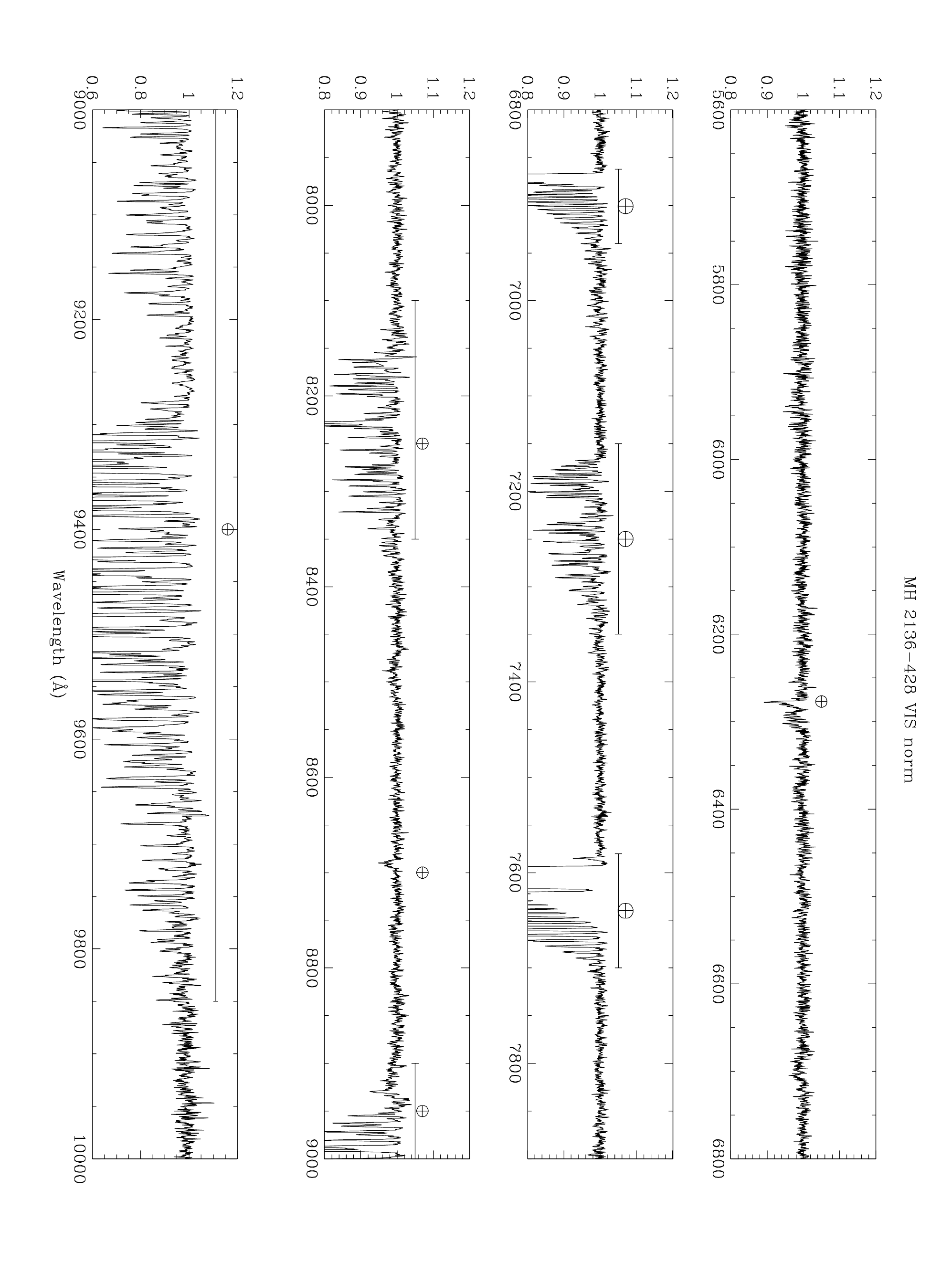}
     \caption{MH 2136-428 X-SHOOTER normalized optical spectrum.}
     \label{fig:2136-visn}
\end{figure*}
\end{center}

\setcounter{figure}{0}
\begin{center}
\begin{figure*}
   \includegraphics[width=18cm]{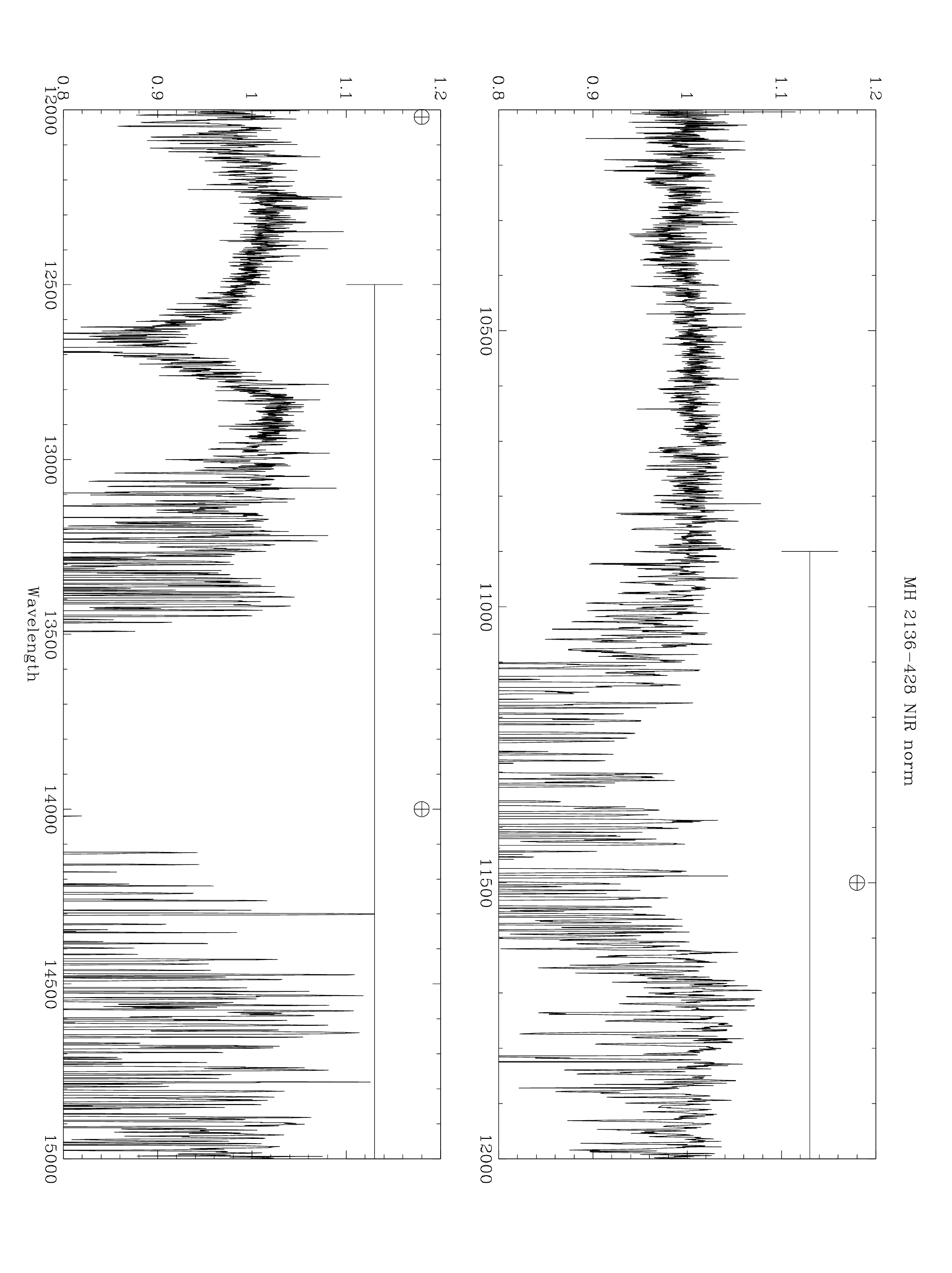}
     \caption{MH 2136-428 X-SHOOTER normalized near-IR spectrum.}
     \label{fig:2136-nirn}
\end{figure*}
\end{center}

\setcounter{figure}{0}
\begin{center}
\begin{figure*}
   \includegraphics[width=18cm]{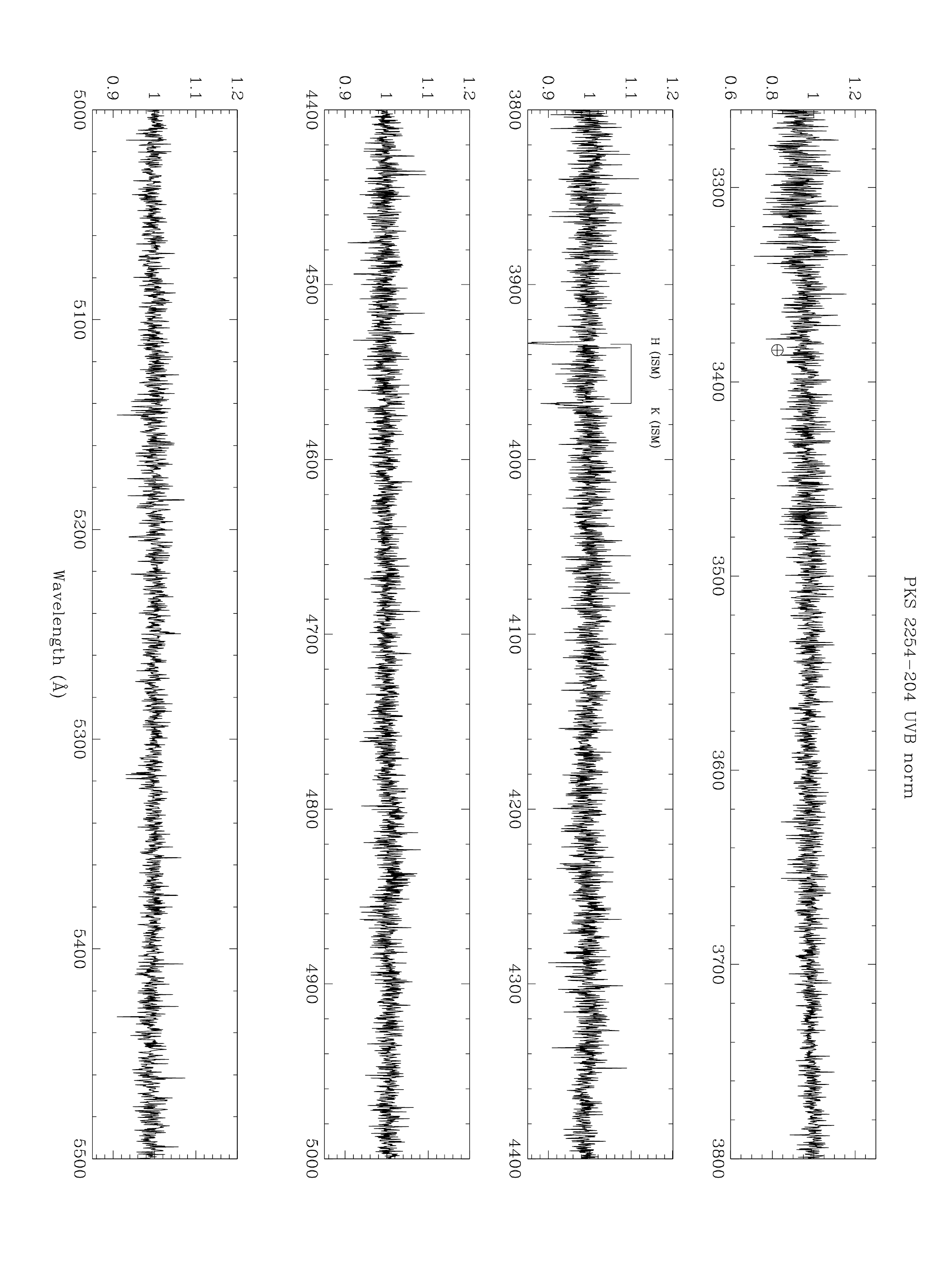}
     \caption{PKS 2254-204 X-SHOOTER normalized UVB spectrum.}
     \label{fig:2254-uvbn}
\end{figure*}
\end{center}

\setcounter{figure}{0}
\begin{center}
\begin{figure*}
   \includegraphics[width=18cm]{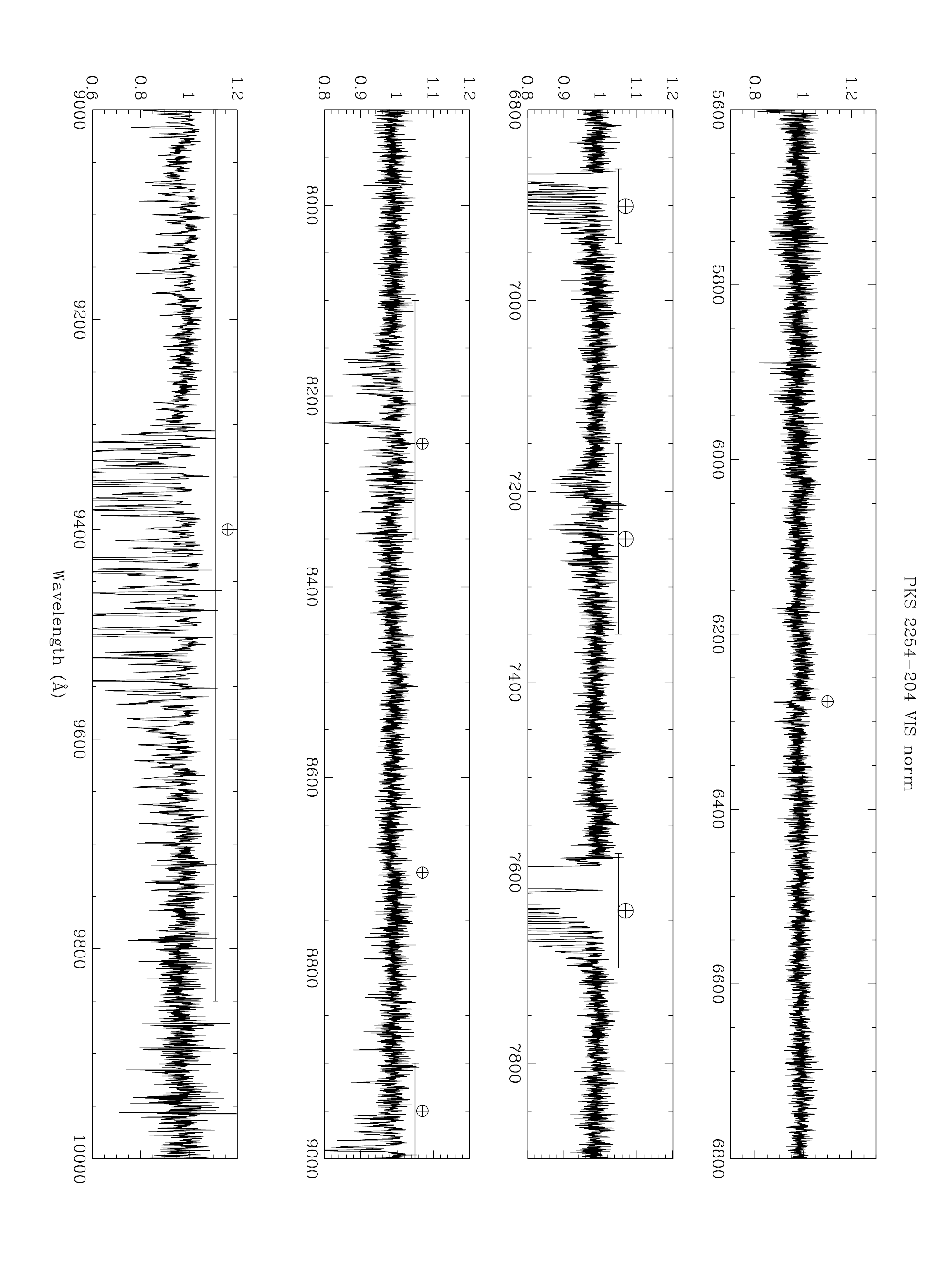}
     \caption{PKS 2254-204 X-SHOOTER normalized optical spectrum.}
     \label{fig:2254-visn}
\end{figure*}
\end{center}

\setcounter{figure}{0}
\begin{center}
\begin{figure*}
   \includegraphics[width=18cm]{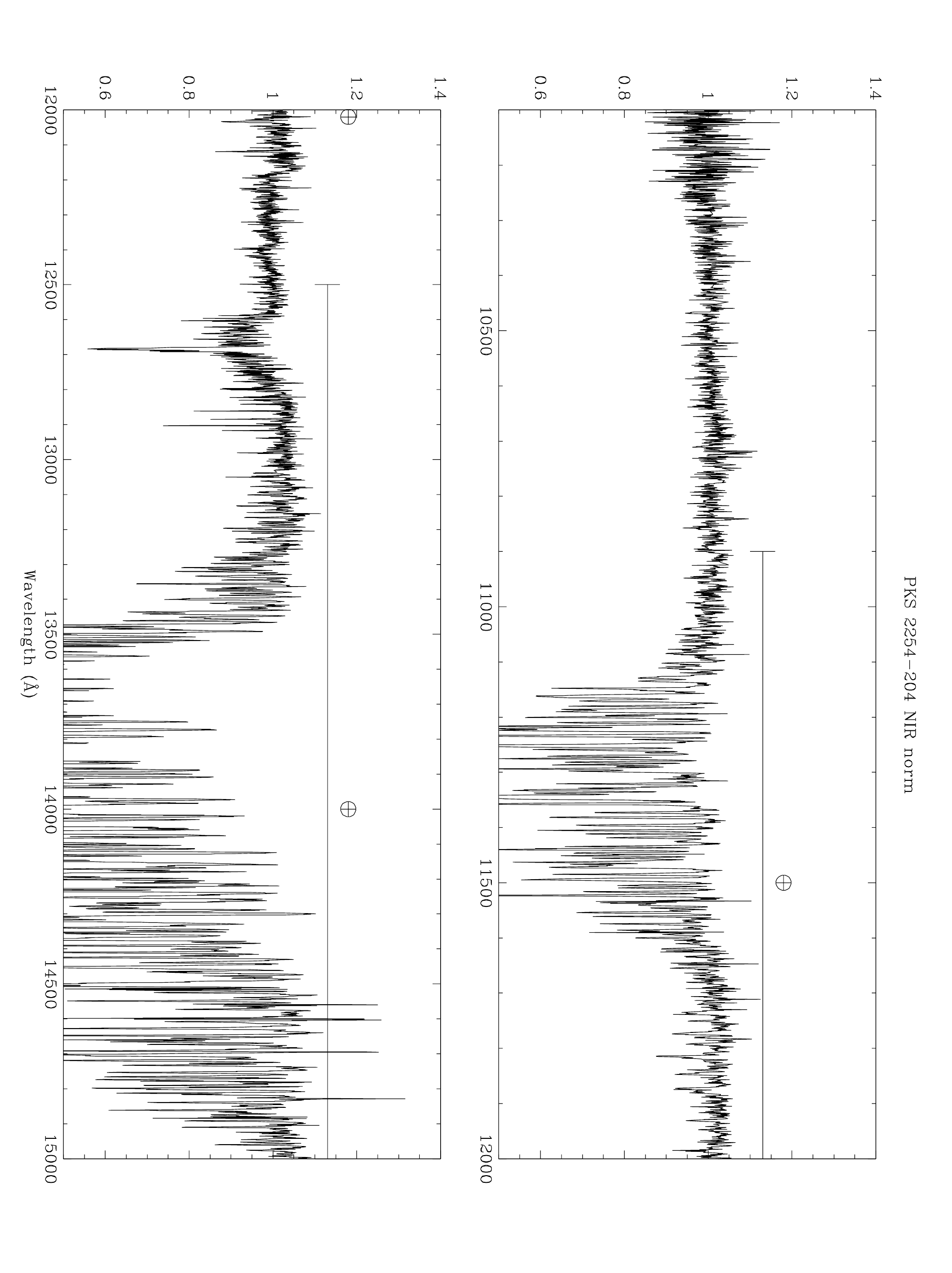}
     \caption{PKS 2254-204 X-SHOOTER normalized near-IR spectrum.}
     \label{fig:2254-nirn}
\end{figure*}
\end{center}

\end{appendix}
\bibliographystyle{aa} 

\bibliography{biblio}{}
 \nocite{*}

\end{document}